# EARTH OBSERVATION AND THE NEW AFRICAN RURAL DATASCAPES: DEFINING AN AGENDA FOR CRITICAL RESEARCH

Rose Pritchard, Global Development Institute, University of Manchester, rose.pritchard@manchester.ac.uk

Wilhelm Kiwango, University of Dodoma, wiaki2007@gmail.com

Andy Challinor, School of Earth and Environment, University of Leeds, a.j.challinor@leeds.ac.uk

**Abstract:** The increasing availability of Earth Observation data could transform the use and governance of African rural landscapes, with major implications for the livelihoods and wellbeing of people living in those landscapes. Recent years have seen a rapid increase in the development of EO data applications targeted at stakeholders in African agricultural systems. But there is still relatively little critical scholarship questioning how EO data are accessed, presented, disseminated and used in different socio-political contexts, or of whether this increases or decreases the wellbeing of poorer and marginalized peoples. We highlight three neglected areas in existing EO-for-development research: (i) the imaginaries of 'ideal' future landscapes informing deployments of EO data; (ii) how power relationships in larger EO-for-development networks shape the distribution of costs and benefits; and (iii) how these larger-scale political dynamics interact with local-scale inequalities to influence the resilience of marginalised peoples. We then propose a framework for critical EO-for-development research drawing on recent thinking in critical data studies, ICT4D and political ecology.

**Keywords:** Earth Observation; agriculture; power; inequality; Africa

## 1. INTRODUCTION

Recent years have seen a rapid increase in the availability of Earth Observation (EO) data, derived both via remote sensing and via ground-based sensors such as weather stations, river flow gauges and camera traps (Gabrys, 2016; Bakker & Ritts, 2018). Coupled with increased computational power, rapidly evolving analytical methods, and the increasing prevalence of ICTs to facilitate data analysis and information dissemination, these data are transforming monitoring and predictive capacities in global land systems (Bakker & Ritts, 2018; Lioutas et al., 2020).

Data availability has been highlighted as a barrier to sustainable development in African countries (Espey, 2019) and EO data can play a useful role in addressing data gaps. Specifically in the context of agriculture, a growing number of applications based on EO data are being developed to inform 'data-driven' agricultural policy and practice in African countries (Bégué et al., 2020; Nakalembe et al., 2021), from online platforms for national-scale decision-makers to mobile phone apps for individual farmers. EO data are celebrated as facilitating 'better' decisions which will lead to greater wellbeing and resilience among individuals and communities, particularly as climate change increases levels of uncertainty in agricultural systems (Jones et al., 2015; Dinku, 2020).

Questions remain, however, over the extent to which these claims are being realised. Technical methods papers vastly outweigh critical social science scholarship on datafication of rural landscapes in African countries (Adams, 2019; Rotz et al., 2019; Klerkx et al., 2019). Few studies question the politics of EO data themselves or pose critical questions over who can access, create, share, use or benefit from EO-derived information in different socio-political contexts (Taylor, 2017; Bakker & Ritts, 2018; Gabrys, 2020). Echoing a point made by Heeks and Shekhar (2019) regarding big data and development, and drawing a parallel with Shelton et al.'s (2015) work on Smart Cities, the academic literature contains many optimistic visions of what EO data *could* be used to achieve, but provides a patchy view of who gains or loses in 'actually existing' datafied agricultural systems.

Here we argue for more critical research into how increasing EO data availability could reshape rural landscapes and rural wellbeing in African countries, focusing particularly on agricultural applications. We begin by highlighting key issues underserved in current EO-for-development literature. We then outline a conceptual approach drawing on research in critical data studies, ICT4D and political ecology.

## 2. KEY OVERSIGHTS IN EO-FOR-DEVELOPMENT RESEARCH IN AFRICAN COUNTRIES

EO data can underpin diverse kinds of agriculture-related information, with examples including weather and climate forecasts, pest early warning and land degradation monitoring (Alexandridis et al., 2020). This information may be useful in itself or combined with other data sources to produce services such as famine early warning or index-based insurance (Baudoin et al., 2016; Ntukamazina et al., 2017). EO-derived information is being disseminated via diverse routes depending on the target users, including online platforms, bulletins, mobile phone apps, agricultural extension systems, and radio broadcasts (Hudson et al., 2017; Munthali et al., 2018).

We focus this piece on African rural contexts because decision-makers in many African countries have historically faced major challenges accessing quality 'data-for-development' (Jerven, 2013). EO, particularly satellite remote sensing, means that certain kinds of data are now becoming available at spatial and temporal resolutions which would have been logistically impossible with 'traditional' survey methods alone. This has triggered a rush of EO-for-development efforts focused on African agriculture, with information products targeted at stakeholders including government agencies, agribusinesses, non-governmental organisations and individual smallholder farmers.

But critical academic research on the datafication of agriculture is still skewed towards European and North American contexts, with less critical research in African countries (Rotz et al., 2019; Klerkx et al., 2019). This could be for a multitude of reasons, such as the comparative recency of the EO-for-development trend, challenges of access when some forms of EO data are privately owned, the continued dominance of 'technical' scientific knowledge in EO-for-development initiatives, or the fact that constrained project timelines leave little space for rigorous impact assessment (Tall et al., 2018). Whatever the cause, this results in important omissions in academic EO-for-development literature focused on African contexts. Here we outline three such omissions, while recognising that these are unlikely to be the only gaps. While we talk generally at time about 'African rural contexts', we recognise that there will be huge variation between and within African countries – which only makes the questions we pose here even more important.

### 2.1. Which development discourses shape the use of EO data?

Our first question is over the discourses of development shaping the ways that EO data are analysed, packaged, disseminated and used. African rural landscapes are contested spaces with many different possible futures, and this makes it essential to question the assumptions and beliefs over 'ideal' landscape futures which inform the design of EO data applications.

This recalls debates over digital agriculture and precision agriculture (summarised in Lajoie-O'Malley et al., 2020). Both of these have been critiqued as embedded in intensive agricultural paradigms which prioritise increased production while downplaying environmental harms. Proponents highlight the social goods arising from intensified agriculture over the last century as well as the benefits of increased production to individual farmers. Critics, in contrast, point out that increased production does not necessarily translate to greater food security or wellbeing, and argue that productivist approaches are legitimised by flawed Neo-Malthusian claims around population growth and resource scarcity (Klerkx & Rose, 2020; Lajoie-O'Malley et al., 2020).

Similar productivist narratives – that EO-derived information can increase efficiency and therefore yields, so ensuring food security for a growing population – are clearly apparent in the rhetoric around EO data services in Africa (see e.g. Dinku, 2020). Less certain is the extent to which EO data are being operationalised to support the implementation of alternative agricultural visions, such as those centring principles of agroecology, food justice or food sovereignty (the latter discussed by Fraser, 2020).

### 2.2. How do power relationships shape – or become reshaped by – access to and use of EO data?

These different future imaginaries cannot be considered in isolation from the relationships of power structuring the networks involved in producing, analysing, disseminating and using EO data. The EO-for-development scene in African countries is messy and fragmented, involving complex assemblages of actors from both global North and South; Blundo-Canto et al. (2021) identified 161 organisations involved in scaling up climate services for Senegal alone. The changes in agricultural systems arising from EO data availability, and the distribution of costs and benefits, will be determined by the power of different actors to influence how EO data are accessed, packaged and used.

The changing availability and value of EO data will alter these power relationships in turn. In some cases, this could compound existing asymmetries, as in the case of precision agriculture and large agribusiness (Lioutas et al., 2020). In others it could create new asymmetries, as in the example of commercially valuable data held by meteorological departments (Nordling, 2019). And in some this could allow asymmetries to be challenged, as can be the case in counter-mapping initiatives (Peluso, 1995; although see Wainwright & Bryan, 2009).

Despite this, few studies explore power relationships in larger EO-for-development networks or consider the distribution of benefits and costs. This is true even in the case of weather and climate services, which are the longest-standing and thus best-studied kind of EO data service. Harvey et al. (2019) offer a rare example exploring the roles of NGOs in climate service delivery, and conclude with a call for greater research into the politics and power dynamics of climate service networks. Vogel et al. (2019) reach a similar conclusion that the political economy of climate services has been largely neglected. Several authors have now discussed the role of power dynamics in shaping climate service co-development approaches (e.g. Daly & Dilling, 2019; Vincent et al., 2020), but these studies are often reflections on individual projects rather than explorations of extended networks.

### 2.3. Who benefits from EO-derived information at local scales?

An incomplete engagement with how the costs and benefits of EO data applications are distributed is also apparent at local scales. A small number of studies have explored how characteristics such as age, gender or income shape ability to benefit from EO-derived information in African rural contexts, documenting access inequalities which will be familiar from the wider ICT4D literature (Muema et al., 2018; Gumucio et al., 2020). But as Nyantakyi-Frimpong (2019) observes, there are few studies in this literature which adopt an intersectional lens, despite the intersections of multiple characteristics being so important for shaping vulnerability and adaptive capacities (Turner et al., 2003; Erwin et al., 2021).

Current literature on local impacts of EO-derived information also emphasises benefits without much consideration of potential risks (echoing observations by Clarke, 2016; Barret and Rose, 2020).

Writing specifically on climate services, Nkiaka et al. (2019) found a substantial number of case studies focusing on individual-scale changes in yields and incomes, but there are less if any studies considering emergent impacts on community-scale dynamics in African contexts.

This is an important gap because it is widely recognised that the direct benefits of agricultural information services are less like to accrue to the most vulnerable people in communities (Lemos & Dilling, 2007; Roudier et al., 2016). It remains unknown whether altered behaviour among the more advantaged will produce co-benefits, or whether it will further marginalise and disempower more vulnerable community members.

## 3. DEFINING A CRITICAL EO-FOR-DEVELOPMENT RESEARCH AGENDA

These three under-researched areas call to mind recent work by Eriksen et al. (2021), who found that adaptation interventions designed without reference to patterns of power and inequality often result in maladaptive outcomes for poorer and marginalized peoples. Now is a good time to evaluate whether these same issues are being reproduced in EO-for-development efforts, or conversely whether EO data applications are delivering the promised benefits. We propose to explore this using a conceptual framework drawing on research in critical data studies, political ecology and ICT4D (summarized in Figure 1).

We begin by characterizing EO data as forming landscape 'data doubles' – a term developed initially in surveillance research (Haggerty & Ericson, 2000), but here adapted to refer to the abstracted version of rural landscapes created through EO data. This full suite of available data goes through a series of filtering processes, firstly being reduced to what we term a 'datascape' – a simplified spatial representation of the landscape. The data double and datascape are neither neutral nor complete. As shown by research in critical cartography (e.g. Harris & Hazen, 2005), they are a function of the data available (which is itself politically determined) and the priorities of those creating the representation. But as Venot et al. (2021) demonstrate in their recent work on irrigation data, even flawed representations can have substantial influence.

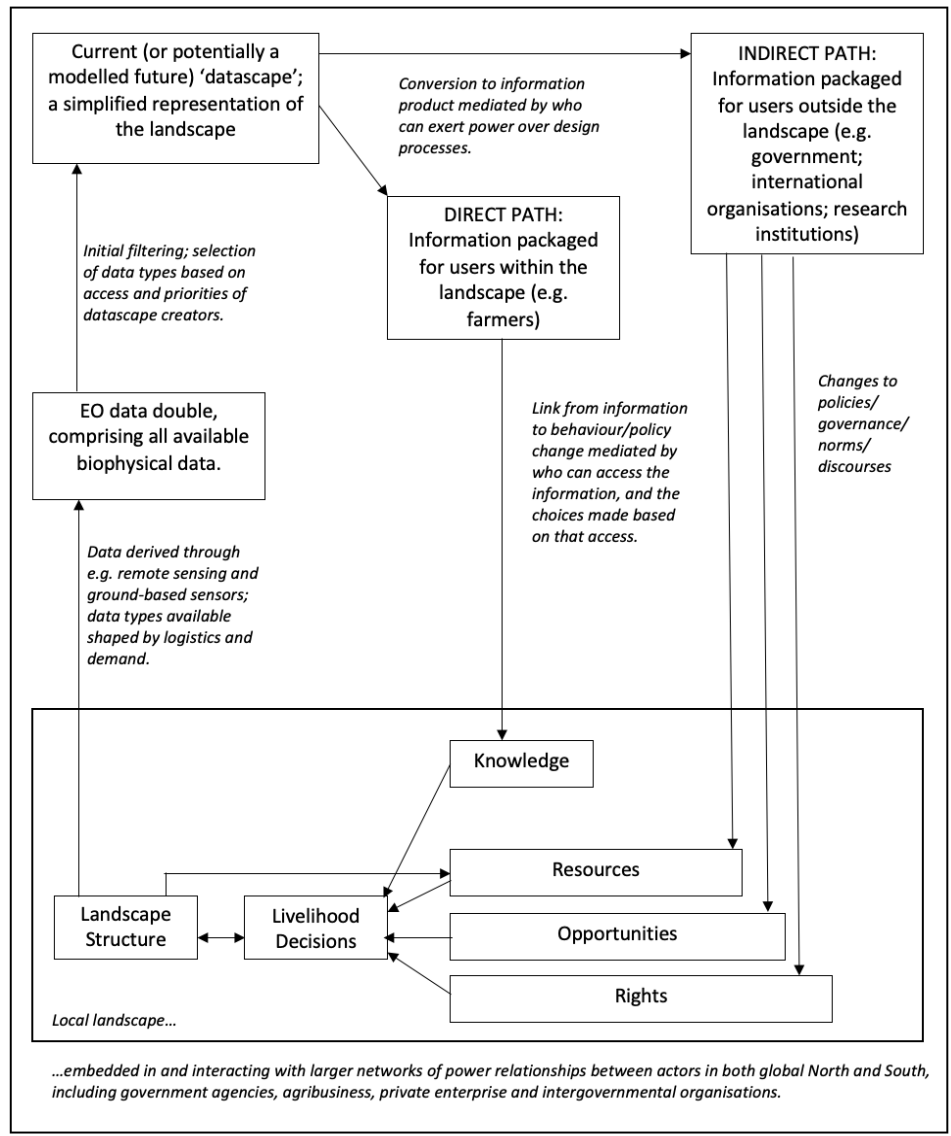

**Figure 1 A conceptual framework for exploring how increasing Earth Observation data availability could reshape land and livelihoods in rural African landscapes. We situate the local landscape within larger-scale networks of power relationships, which will determine the values and priorities shaping the production, packaging, dissemination and use of EO data. This in turn will influence the nature and distribution of costs and benefits both within rural landscapes and through the larger network.**

These datascapes often form the basis for a range of information products. These products may be targeted directly at land managers such as farmers or may impact land and livelihoods indirectly by influencing policy and governance. Here theory from critical data studies is essential, emphasizing as it does the importance of power dynamics and political processes of negotiation and contestation among complex networks of actors (boyd & Crawford, 2012; Dalton & Thatcher, 2014; Jasanoff, 2017). Similar discussions of power and participation are common in the ICT4D literature (Daly & Dilling, 2019), which has particular relevance given that many information products are disseminated via online platform or mobile phone. Which actors are able to exert power over decision-making processes, and the worldviews and values of these actors, will shape the nature and distribution of costs and benefits both at local scales and through the larger network.

Much of the research on Big Data focuses on the social impacts of datafication. We also propose to draw on political ecology, which shares similar themes to critical data studies in terms of emphasizing dynamics of power, but places more explicit focus on changes to land as well as livelihoods. There is a rich tradition of research in political ecology on how different views of 'ideal' rural landscapes come be to seen as valid and legitimate while others are sidelined, and how this can lead to both social and ecological harms (e.g. Fairhead & Leach, 1996; Asiyanbi, 2016). Research drawing on both critical data studies and political ecology is already providing useful insights in the field of natural resource governance. McCarthy and Thatcher (2019) use theories from both to explore the construction and potential impacts of World Bank resource maps, while Iordachescu (2021) discusses how the ways that Romanian landscapes are characterized based on remote sensing erases local people from landscape histories.

The kind of complex systems research proposed here poses interesting methodological challenges. One option is to focus on particular strands of EO data and track them through networks, as in Bates et al's (2016) 'data journeys' approach. Another is to begin with impacts at the local landscape scale, as in the case studies reviewed by Nkiaka et al. (2019), and seek to reconstruct the processes leading to particular outcomes. A third is to focus on a particular stage in the data filtering process, as in the growing body of literature on the co-production of climate services (Vincent et al., 2018; Daly & Dilling, 2019). In practice, a combination of these approaches is likely to yield the greatest insights – particularly because the networks we reference here and the power relationships within them are dynamic, interacting with EO data sources which are also constantly evolving.

## 4. CONCLUSION

Our objective in this piece was to highlight the need for more critical research on how the growing ubiquity of EO data is reshaping African rural landscapes. EO data do open up exciting opportunities in data-sparse contexts, but equally raise new questions and challenges – particularly with regards to how the increasing availability of these data will interact with complex power relationships in EO-for-development networks and the consequences this will have for the people living in rural landscapes. We believe that research drawing on critical data studies and political ecology could provide valuable insights into the distribution of costs and benefits arising from EO data, and thereby help identify ways of realizing the potentials of EO data while minimizing the possible harms.

## REFERENCES AND CITATIONS